\documentclass[twocolumn,amsmath,amssymb]{revtex4}

\usepackage{amsmath,amssymb}
\usepackage{graphicx}
\usepackage{epsfig}
\usepackage{dcolumn}
\usepackage{bm}

\begin{document}
\title{Cooperative reliable response from sloppy gene-expression dynamics}
\author{Masayo Inoue$^1$ and Kunihiko Kaneko$^2$}
\affiliation{
$^1$School of Interdisciplinary Mathematical Sciences, Meiji University, 4-21-1 Nakano, Nakano-ku, Tokyo 164-8525, Japan \\
$^2$Research Center for Complex Systems Biology, Universal Biology Institute, University of Tokyo, 3-8-1 Komaba, Tokyo 153-8902, Japan
}
\begin{abstract}
Gene expression dynamics satisfying given input-output relationships were investigated by evolving the networks for an optimal response. 
We found three types of networks and corresponding dynamics, depending on the sensitivity of gene expression dynamics: direct response with straight paths, amplified response by a feed-forward network, and cooperative response with a complex network. 
When the sensitivity of each gene's response is low and expression dynamics is sloppy, the last type is selected, in which many genes respond collectively to inputs, with local-excitation and global-inhibition structures. 
The result provides an insight into how a reliable response is achieved with unreliable units, and on why complex networks with many genes are adopted in cells.
\end{abstract}
\maketitle

Information processing based on the on$/$off behaviors of units is ubiquitous and essential in biological systems, such as in neural and gene regulatory systems. 
The response of each unit is not as reliable as digital units in computers and often shows a sloppy response. 
Then, the question arises as to how reliable information processing can be achieved with such sloppy units. 
Indeed, in his pioneering publication, von Neumann addressed such a question, with reliable computation from unreliable units, and proposed the majority rule by averaging the outputs of multiple unreliable units \cite{Neumann:1956aa}.
 
This question is not restricted to computation by unreliable electronic elements or neurons. 
On$/$off behaviors are also common in gene expressions, from which cellular outputs depending upon inputs are generated. 
These expression dynamics are also not digital. 
The sensitivity of each expression is gentle compared to the step function. 
Indeed, the Hill coefficient $n$ representing this sensitivity is typically $2 \sim 4$ \cite{Becskei:2005aa, Rosenfeld:2005aa, Dekel:2005aa, Kim:2008aa} (the step function is realized for $n \rightarrow \infty$). 
Then, how reliable output is generated from such sloppy gene expressions has to be explored.

To explore the cellular input-output (I/O) behavior of gene expression patterns or other biochemical reactions, network analysis is often adopted. 
In particular, the roles of simple network motifs with a few nodes are widely identified.  
Considering simple on$/$off units and network motifs with a few nodes, appropriate I/O behaviors can be designed as in logical circuits.  
Indeed, such architecture is sometimes observed in biological networks \cite{Alon:2006aa, Ma:2009aa}.
If the units are digital, the desired output can be designed by simply combining the motifs.  
However, the question remains how reliable I/O behaviors are generated when sloppy units are adopted. 
 
Note that, in most real biological systems, the network structure is not as simple as expected from a series of network motifs.
Paths in the gene regulatory network (GRN) are intermingled, and independent motifs are difficult to extract.
For example, studies using DNA microarrays of yeast \textit{Saccharomyces cerevisiae} have shown that more than half of the genes in GRN respond to every environmental changes \cite{Gasch:2000aa, Causton:2001aa, Gasch:2002aa}. 
Moreover, their responses are often continuous, not digital, between on$/$off states.
Several studies have shown that many genes (i.e., 50\%--70\%) exhibit adaptive responses (i.e. up-down or down-up transient response) with respect to the inputs \cite{Deutscher:2006aa, Stern:2007aa, Furusawa:2012aa}.
These observations cannot be explained with a combination of motifs constituting logic circuits.
 
Here we explore how expression dynamics by a regulatory network of multiple genes shape appropriate I/O relationships.
By using a genetic algorithm for network selection, we uncover three distinct types of dynamics to achieve a proper I/O relationship: direct, feed-forward, and cooperative networks. 
In the cooperative type, a reliable I/O relationship is generated from units with low Hill coefficients, and local excitation and global inhibition (LEGI) are revealed as its characteristic behavior.

\begin{figure}[b]
\includegraphics[width=85mm]{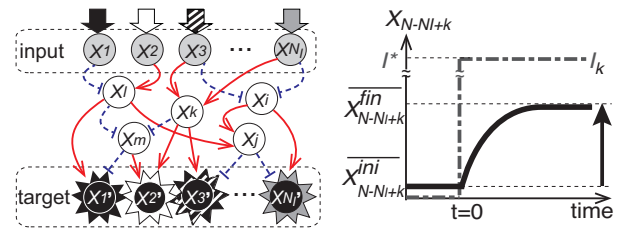}
\caption{(left) Schematic view of the GRN model. Each circle represents a gene with expression level $x_i$ and arrows with solid red lines (broken blue lines) show excitatory (inhibitory) interactions. The target gene ($k'=N-N_T+k$) should respond upon the external input on the input gene $k$ specifically. (right) Definition of our evaluation function. The input ($I_k$) is shown by the broken line, and the response of the corresponding target gene is shown by the solid line. $\overline{x_i^{ini}}$ and $\overline{x_i^{fin}}$ are defined as the average values over some time span. 
  \label{fig_model_network}}
\end{figure}

We adopt a simplified GRN model \cite{GLASS:1973aa, MJOLSNESS:1991aa, Salazar-Ciudad:2001aa, Kaneko:2007aa, Furusawa:2008aa, Inoue:2013aa}. 
It is composed of $N$ genes as nodes in the network, which are divided into three types: $N_{I}$ input genes receiving external inputs, $N_{T}$ target genes providing the output and determining the fitness of the cell, and $N_{M}$ middle-layer genes (ML-genes) that transmit the input to the target ($N_{I} + N_{T} + N_{M} = N$).
We consider one-to-one correspondences between an input and a target gene, so that we set $N_{I} = N_{T}$ by assigning the nodes $1, \dots, N_I$ as the input genes, $N-N_I+1, \dots, N$ as the target genes, while leaving others as ML-genes (Fig.\ref{fig_model_network}). 
Below, we first present the result for $N=100$, $N_{I} = 5$ and $N_{M}=90$, and show the generality of the result later. 

Through suitable normalization, the expression level of a gene is represented by a variable $x_i =[0,1] (i=1, \dots, N)$, with the maximal expression level scaled to unity.
The time evolution of the expression is given as follows:  
\begin{eqnarray}
   \frac{dx_i}{dt} &=& \frac{1}{1+\exp \left[-\beta (y_i - y_T)\right]} - \gamma x_i, \label{eq_dynamics} \\
   \gamma &=& 1+\frac{1}{N_{T}} \sum_{j=1}^{N_T} x_{N-N_T+j}.  \label{eq_gamma}
\end{eqnarray}
The first term in Eq.(\ref{eq_dynamics}) represents interactions with other genes, where $y_i = I_k \delta_{ik} + \sum_{j=1}^{N} C_{ij} x_j$ is the total signal that the $i_{th}$ gene receives with $\delta_{ik}$ as the Kronecker delta for $k=1, \dots, N_I$. 
$I_k$ shows the external input on the $k_{th}$ input gene. 
$C_{ij}$ represents the regulation from gene $j$ to $i$, with $1$ (excitatory), $-1$ (inhibitory), and $0$ (non-existent). 
Here the input genes do not receive regulation from others ($C_{ij}=0$ for $i=1, \dots, N_I$) and the target genes do not regulate others ($C_{ji}=0$ for $i=N-N_T+1, \dots, N$).  
$y_T$ denotes a constant threshold and $\beta$ determines response sensitivity corresponding to the Hill coefficient. 
For simplicity, we assume that all genes in a network have the same $y_T$ and $\beta$ values. 
As $\beta$ becomes larger, the first term approaches a step function with a threshold $y_T$. 
The second term represents degradation and we assume that the degradation depends on the total expression levels of the target genes (Eq.(\ref{eq_gamma})) (see below).  

Initially, the expression level of each gene is set to a randomly chosen level between $0$ and $1$, and evolves according to Eq.(\ref{eq_dynamics}) with $I_k=0 (k=1, \dots, N_I)$, until $x_i$ reaches a steady state. 
Then, the external input on a single input gene is applied at $t=0$, by switching to $I_{l}=I^{\ast}, I_j =0 (j\neq l)$. 
$I^{\ast}$ is set to $5$, whereas the results below are not affected as long as $I^{\ast} \gg y_T$. 

For evolution, the paths in the regulation matrix $C_{ij}$ are mutated and such a $C_{ij}$ is selected according to the following fitness condition: 
We assume that the target gene $N-N_T+k$ should respond following the application of $I_k$ ($k=1, \dots, N_I$), and the fitness is defined as the average of responses against each $I_k$.
It is given as the difference between the final and initial expression levels of the corresponding target as
\begin{eqnarray}
   fitness = \frac{1}{N_I} \sum_{k=1}^{N_I} (\overline{x_{N-N_T+k}^{fin}}- \overline{x_{N-N_T+k}^{ini}}). 
 \label{eq_fitness}
\end{eqnarray} 
where $\overline{x_j^{fin}}$ is the temporal average of $x_j$ between $T_1 < t < T_2$ for sufficiently large $T_1$ and $T_2$ and $\overline{x_j^{ini}}$ is the average between $-T_0 < t <0$. 

Note that Eq.(\ref{eq_fitness}) might allow for the trivial solution in which all target genes respond to any input, rather than a one-to-one response.
To eliminate such a possibility, and also to take into account the cost of the expression, a punishment term is included in the definition of $\gamma$ in Eq.(\ref{eq_gamma}), so that the expression of target genes will give the dilution of each expression $x_i$ (one could also interpret that the cell volume increases in proportion to the expression of proteins). 
Hence, expressing all target genes will result in a decrease in the fitness. 

For the next generation, the network structure, i.e. the regulation matrix $C_{ij}$, is slightly modified by "mutation", whereas the parameters are kept unchanged. 
In the mutation process, we fix the number of paths and swap the connection with a small mutation rate ($1$ or $3$ paths are mutated on average for every process).
For each generation, $100$ networks are prepared and $25$ networks with the highest fitness (Eq.(\ref{eq_fitness})) are selected.
From these, $100/25 = 4$ mutant networks are generated and the selection process is repeated with the $100$ newly generated networks as a simple genetic algorithm.

After the evolution, the highest fitness with one-to-one correspondence between input-target pairs is achieved, regardless of $\beta$ and $y_T$ values. 
In an ideal situation, only the corresponding target gene responds by changing from $\overline{x_i^{ini}} =0$ to $\overline{x_i^{fin}} =x^{\ast}$. 
Then, the steady-state solution in Eq.(\ref{eq_dynamics}) with $\gamma = 1 + x^{\ast} / N_{I}$ leads to $x^{\ast} = (-N_{I} + \sqrt{N_{I} (N_{I} + 4)}) /2$ which is $\approx 0.85$ in the case of $N_{I} = 5$. 
The maximal fitness value of Eq.(\ref{eq_fitness}) is also given by $x^{\ast}$. 

\begin{figure}[b]
\includegraphics[width=85mm]{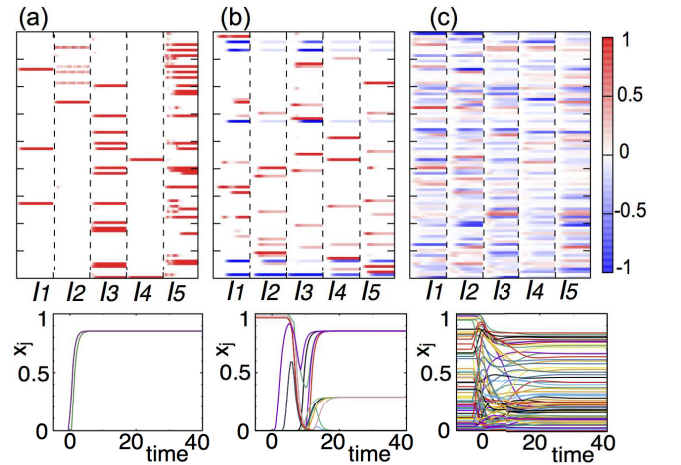} 
\caption{Three types of behaviors of the ML-genes. The fittest networks with (a) $\beta = 100, y_T = 0.5$, (b) $\beta = 10^{1.5}, y_T = 0.875$, (c) $\beta = \sqrt{10}, y_T = 0.25$. 
(upper) $x_j (t) - \overline{x_j^{ini}}$ at $t=0, 1, \dots, 20$ as the abscissa is plotted by using the color map, with $j$ as the vertical axis, for the application of each $I_k$ ($k=1, \dots, 5$). 
(lower) Temporal changes in the expression levels of all $90$ ML-genes are overlaid when $I_1$ is applied. 
  \label{fig_3dynamics}}
\end{figure}

The responses of the ML-genes do not affect the fitness function in Eq.(\ref{eq_fitness}) at all. 
However, according to their behaviors, three distinct types of the evolved dynamics are uncovered, depending on $\beta$ and $y_T$ values. 
In the first type that appears for a large $\beta$ and intermediate $y_T$, only a small number of ML-genes show monotonic increase. 
Each ML-gene responds to specific, usually only one, external inputs and remains unchanged for other inputs (Fig.\ref{fig_3dynamics} (a)). 
In the second type for large $\beta$ and large $y_T$, the number of responding ML-genes increases, but they still respond only to each specific input (Fig.\ref{fig_3dynamics} (b)). 
Some show monotonic increase or decrease, and few others show non-monotonic, adaptive responses between on- ($x_i \sim 1$) and off- ($x_i \sim 0$) states. 
Many (more than half) ML-genes do not respond to any inputs, even if they are connected to responding ML-genes, due to inhibitory regulations from others.  
In the third type, unlike the previous two types, almost all ML-genes respond whenever any external inputs occur.  
This type appears for small $\beta$ or small $y_T$ values. Each of the ML-genes shows different responses to different inputs. 
Not only monotonic but also adaptive responses are observed in both increasing and decreasing directions (Fig.\ref{fig_3dynamics} (c)). 

\begin{figure}[tb]
\includegraphics[width=85mm]{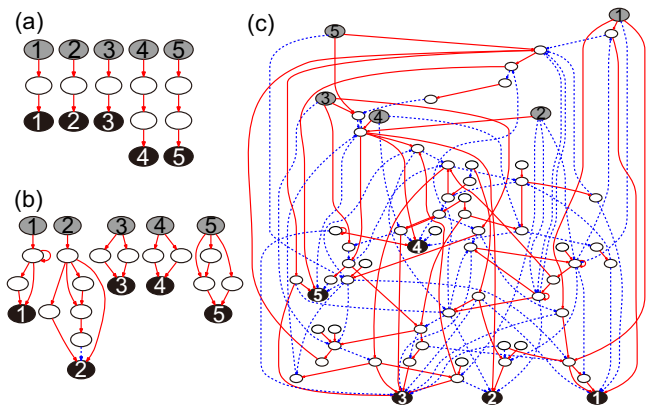}
\caption{Typical example of each type of core structure. See text for the definition of a core structure. Gray and black circles show input and output genes, respectively, and the numbers indicate their correspondences. Red bold arrows represent excitatory regulations and blue dotted arrows denote inhibitory regulations. The same networks with Fig.\ref{fig_3dynamics} are used. 
  \label{fig_3typenetwork}}
\end{figure}

These three types show different characteristics in the network structure. 
Results from network motif analysis are shown in Supplemental Material $1$. 
However, the differences in the network structure are clearer when compared with a core structure; the core structure is obtained by removing a path randomly, one-by-one, as long as the corresponding target gene response to each input is preserved, even if it may be a bit lower (this condition is given by keeping $fitness > 0.7$) (Fig.\ref{fig_3typenetwork}). 

The core structure of the first type simply connects an input gene and the corresponding target gene independently with a straight excitatory interaction (Direct type; Fig.\ref{fig_3typenetwork} (a)). 
Each input-target pair is connected via one or a few ML-genes. 
This is the type we can easily design as the fittest network with the current fitness condition, for digital units with large $\beta$. 
In the second type, a feed-forward (FF) network structure that independently connects each input gene to the corresponding target pair is formed. 
The external input is amplified with this FF structure, which is relevant to a unit with larger $y_T$ (FF-network type; Fig.\ref{fig_3typenetwork} (b)).  
The third type is not as simple as the previous two types and contains many ML-genes connected to each other, both with excitatory and inhibitory interactions (Cooperative type; Fig.\ref{fig_3typenetwork} (c)). 
No input-target pair is independent and most ML-genes are shared by several pairs.  

\begin{figure}[tb]
\includegraphics[width=85mm]{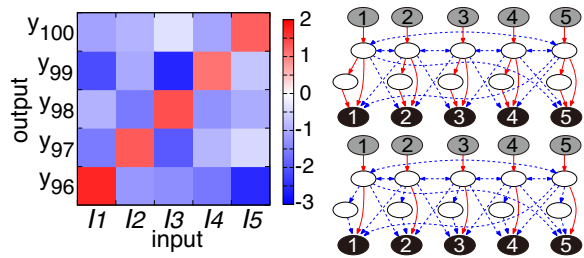}
\caption{(left) The correspondence relation between input and target genes in the cooperative type. The $y_i$ values of target genes (ordinate) when each $I_k$ ($k=1, \dots, 5$) is applied (abscissa) are shown. Each target gene receives an excitatory regulation in the case of the corresponding $I_k$ and an inhibitory regulation on other occasions. 
(right) Simple network examples with the LEGI structure. The fitness values are $0.372$ (top) and $0.506$ (bottom) with $\beta = \sqrt{10}$ and $y_T = 0.25$.
  \label{fig_localactivation}}
\end{figure}

In Fig.\ref{fig_3typenetwork} (c), all input genes are indirectly connected to all target genes and each input gene gives excitatory regulations ($y_i>y_T$) to the single corresponding target gene and inhibitory regulations ($y_i<y_T$) to other targets (LEGI; Fig.\ref{fig_localactivation}). 
In contrast, the core structures of the Direct and the FF-network types are composed only of the local-excitation, without global inhibition. 
However, simple networks artificially designed by local-excitation and homogeneous or random global-inhibitory regulations result in much lower fitness, $\sim 0.5$ at most (Fig.\ref{fig_localactivation}).  
Some delicately balanced inhibitory regulations in the evolved networks are essential for high fitness.

\begin{figure}[tb]
\includegraphics[width=85mm]{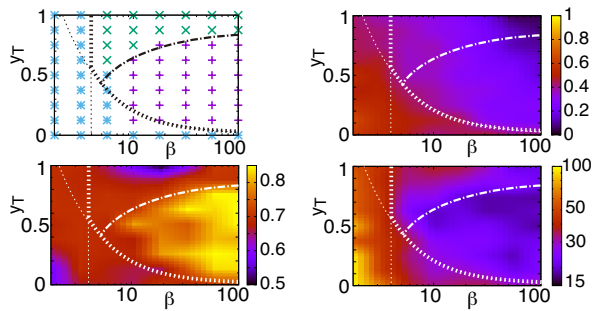}
\caption{Phase diagrams with regard to $\beta$ (abscissa) and $y_T$ (ordinate). Phase diagram based on the core structures (top left); Direct type (purple $+$), FF-network type (green $\times$), and Cooperative type (blue $\ast$).  
The number ratio of responding ML-genes (top right), fitness value of core structures (bottom left), and total number of genes for constructing core structures (bottom right) are shown by using three fittest networks independently evolved with each $\beta$ and $y_T$ value. 
  \label{fig_sozu}}
\end{figure}

The phase diagrams of the three types in terms of $\beta$ and $y_T$ are given in Fig.\ref{fig_sozu}.  
The phase boundaries are estimated as follows: Let $J(y)=1/(1+exp[-\beta(y-y_T)])$ from Eq.(\ref{eq_dynamics}). 
First, if the slope of $J(y)$ at $y=y_T$ that determines the on$/$off sensitivity of each gene is less than unity, a cooperative effect from multiple genes is needed to create an on$/$off response. 
Hence, the non-cooperative types exist for $\frac{dJ}{dy_i} (y_T) > 1$, i.e. for $\beta >4$.
Second, to achieve a high fitness value, $\overline{x_{N-N_T+k}^{ini}} \sim 0$ is needed. 
For a steady-state expression level without an interaction term, this postulates $J(0) \sim 0$ (i.e., for non-cooperative types), whereas inhibitory regulations from ML-genes are necessary if $J(0)\gg 0$ (the Cooperative type). 
Third, $J(x^{\ast}) / \gamma$ gives an expression level of a gene that receives an excitatory regulation from a single highly expressed gene, where $x^{\ast}$ is the maximal steady-state expression level as already defined. 
For the Direct type, high expression just by a single regulation is needed, i.e., $J(x^{\ast}) \sim1$ (or $\gg y_T$), otherwise signal amplifications are necessary ($J(x^{\ast}) \ll 1$; the FF-network type). 

The latter two conditions are estimated by approximating $J(y)$ by $\tilde{J}(y)=0 (y \leqslant y_T - \frac{2}{\beta}), \ \frac{\beta}{4} (y - y_T) +\frac{1}{2} (y_T - \frac{2}{\beta} \leqslant y \leqslant y_T + \frac{2}{\beta}), \ 1 (y_T + \frac{2}{\beta} \leqslant y) $.  
Accordingly, the boundaries are estimated as follows: $\tilde{J}(0) \sim 0$ leads to $y_T = \frac{2}{\beta}$, and $\tilde{J}(x^{\ast})\sim1$ leads to $y_T = x^{\ast} - \frac{2}{\beta}$. 
Hence, besides the line $\beta = 4$, the curve $y_T = \frac{2}{\beta}$ gives a boundary between the Cooperative and other two types, and the curve $y_T = x^{\ast} - \frac{2}{\beta}$ gives that between the Direct and the FF-network types.  
These simple estimations of the phase boundaries, as depicted in Fig.\ref{fig_sozu}, roughly agree with the numerical result. 
Finally, it is interesting to note that around the boundary of the FF-network and the Cooperative types, a mixed network evolves with a core structure combining $5$ feed-forward subnetworks to one (Supplemental Material 2). 

In this letter, we show that the optimal network structure for information processing differs depending on its units' reliability defined by the response sensitivity $\beta$ corresponding to the Hill coefficient and the threshold $y_T$. 
Direct paths connecting an input and a target gene straightforwardly in a line are sufficient when units are reliable (large $\beta$ and intermediate $y_T$; Direct type) whereas FF-network for signal amplification evolve when $y_T$ is larger (FF-network type). 
In these two types, each I/O relationship is achieved independently. 
On another front, networks that achieve the I/O relationships collectively are selected for units with smaller $\beta$ (Cooperative type). 
All target genes are connected to all input genes exhibiting LEGI. 

We have also confined the generality of the three phases we found here, in particular, the Cooperative type for small $\beta$ (Supplemental Material 3). 
First, the three phases exist regardless of $N_I$ and $N_M$, although there exists a lower bound for $N_M$ to achieve the Cooperative type. 
Furthermore, the dependence of the boundary between the Direct and the FF-network types upon $N_I$ is in agreement with the estimate based on $x^{\ast}$.
Second, even if a constant $\gamma$ is adopted instead of Eq.(\ref{eq_gamma}), the three phases are obtained by revising the fitness so that the single corresponding target gene is expressed. 

It is interesting to note that the Cooperative type shows characteristic features common with those observed in biological systems. 
First, a many-to-many correspondence between external inputs and ML-genes; almost all the ML-genes respond to a variety of different inputs. 
Such a relationship has been reported in the expression patterns of yeast \textit{Saccharomyces cerevisiae}. 
Diverse responses far beyond the Direct or FF-network types have been observed \cite{Gasch:2000aa, Causton:2001aa}. 
Moreover, many gene expressions are known to exhibit adaptive, non-monotonic transient responses as a result of complex regulations, as found in our study.

Previously, cooperative adaptive responses in complex GRNs with many genes were revealed to achieve adaptive behavior as outputs \cite{Inoue:2013aa}. 
Here, we found that cooperative responses were relevant, just to create simple I/O relationships with sloppy units. 
The response sensitivity of each unit in our model and in the Hill equation can be related as $n \sim \beta y_T$ \cite{hill}. 
Note that in the gene expression in a cell, the Hill coefficient is typically $2 \sim 4$ \cite{Becskei:2005aa, Rosenfeld:2005aa, Dekel:2005aa, Kim:2008aa}, which corresponds to near the boundary of the cooperative phase in Fig.\ref{fig_sozu}. 
Cooperative response by an intermingled network of many elements will be a general strategy in cellular systems.

The collective and reliable computation with unreliable units was pioneered by von Neumann, where error correction by the simple averaging of such units was adopted \cite{Neumann:1956aa}. 
In contrast, the cooperative response we uncovered here adopts the LEGI network, where the balance between local excitation and global inhibition is a key feature. 
Indeed, such a global inhibition in space was often adopted in biological systems as the global diffusion of inhibitors \cite{Levchenko:2002aa, Takeda:2012aa, Wang:2012aa, Nakajima:2014aa}, whereas the global inhibition in our study is shaped in the network space. 
Detection of the LEGI structure by the global analysis of gene expression patterns and cellular pathways will be important in the future.

Many networks in biological systems are huge and complex, and look redundant for the demanded function. 
It is often pointed out that such redundant networks are relevant in terms of robustness to mutations or noise \cite{Kaneko:2007aa, Wagner:2000aa, Wagner:2007}. 
Our result provides another perspective: achievement of appropriate I/O relationships from unreliable, sloppy units.

MI was supported by Shiseido Female Researcher Science Grant. 
This research was partially supported by a Grant-in-Aid for Scientific Research (S) (15H05746) and Grant-in-Aid for Scientific Research on Innovative Areas (17H06386) from the Ministry of Education, Culture, Sports, Science and Technology (MEXT) of Japan.


\begin{thebibliography}{50}
\bibitem{hill} The sensitivity in input-output relation for $y=f(x)$ can be estimated by $d (logy) / d (logx) = (x/y) (dy/dx)$. The sensitivity at the half-maximal point (i.e., $f(x) =0.5$ for the function $0 \le f(x) \le 1$) is given by $\beta y_T$ for our model $f(x) = 1/(1+\exp \left[-\beta (x - y_T)\right]) $ and $n$ for the Hill equation $f(x) = \theta x^n/(K^n + x^n)$.
\end{thebibliography}
\end{document}